\documentclass[a4paper,10pt]{article}
\usepackage[cp1251]{inputenc}
\usepackage[english]{babel}
\usepackage{amsmath}
\usepackage{amssymb}
\usepackage{graphicx}
\usepackage[pdftex]{color}
\usepackage[square,numbers,comma,sort&compress]{natbib}
\usepackage[pdftex]{hyperref}
\hypersetup{
             final,
             colorlinks=true,
             linkcolor=blue,
             citecolor=red
}

\DeclareMathOperator{\Tr}{Tr}
\DeclareMathOperator{\RE}{Re}
\DeclareMathOperator{\sgn}{sgn}

\begin{document}
\twocolumn[\scriptsize{\slshape ISSN 0021-3640, JETP Letters, 2015, Vol. 102, No. 11, pp. 713--719. \textcopyright\,Pleiades Publishing, Inc., 2015.

Original Russian Text \textcopyright\,P.\,V. Ratnikov, A.\,P. Silin, 2015, published in Pis'ma v Zhurnal Eksperimental'noi i Teoreticheskoi Fiziki, 2015, Vol. 102,
No. 11, pp. 823--829.}

\vspace{0.15cm}

\begin{center}
\rule{7cm}{0.4pt}\hspace{3cm}\rule{7cm}{0.4pt}

\vspace{-0.2cm}
\rule{7cm}{0.4pt}\hspace{3cm}\rule{7cm}{0.4pt}

\normalsize
\vspace{-0.55cm}
{\bf CONDENSED\\
MATTER}

\vspace{0.75cm}

\LARGE{\bf Plasmons in a Planar Graphene Superlattice}

\vspace{0.75cm}

\large{\bf P.\,V. Ratnikov\textit{$^a$} and A.\,P. Silin\textit{$^{a,\,b}$}}

\vspace{0.15cm}

\normalsize

\textit{$^a$Lebedev Physical Institute, Russian Academy of Sciences, Leninskii pr. 53, Moscow, 119991 Russia}

\textit{$^b$Moscow Institute of Physics and Technology (State University),\\
Institutskii per. 9, Dolgoprudnyi, Moscow region, 141700 Russia\\
e-mail: ratnikov@lpi.ru}

\vspace{0.25cm}

Received July 3, 2015; in final form, September 16, 2015
\end{center}

\vspace{0.1cm}
\begin{list}{}
{\rightmargin=1cm\leftmargin=1cm}
\item
\small{Plasmon collective excitations are studied in a planar graphene superlattice formed by periodically alternating regions of gapless graphene and of its gapped modification. The plasmon dispersion law is determined both for the quasi-one-dimensional case (the Fermi level is located within the minigap) and for
the quasi-two-dimensional case (the Fermi level is located within the miniband). The problem concerning
the absorption of modulated electromagnetic radiation at the excitation of plasmons is also considered.}

\vspace{0.05cm}

\normalsize{\bf DOI}: 10.1134/S0021364015230137

\end{list}\vspace{1cm}]

\begin{center}
1. INTRODUCTION
\end{center}

Graphene (a two-dimensional carbon material)
has been actively studied both theoretically and experimentally
for more than ten years. In recent years, graphene nanostructures have become a forefront
issue. The usage of collective excitations (plasmons) in these systems promises new advantages for
the tunable absorption of electromagnetic radiation.
The plasmon-induced enhancement of light absorption
within the middle infrared range was observed for
the heterostructure formed by graphene strips \citep{Freitag}.

The plasmon-type oscillations in spatially uniform
systems with different dimensionalities having charge
carriers with a linear dispersion law were studied in
\citep{DasSarma}, where the tunneling of charge carriers was
neglected. Such approximation is similar to the tight-binding
approximation in the band structure theory
for crystals.

In \citep{Chaplik}, the plasma oscillations of massless Dirac
electrons in a planar superlattice were studied. The
Dirac plasma was assumed to be weakly modulated.
This picture is similar to the weak-binding approximation.
The spectrum of plasma oscillations and the
related absorption intensity for electromagnetic waves
were determined by the methods of electrodynamics
of continuous media.

In this paper, we present the calculations of the
plasmon dispersion law in planar graphene superlattices.
The superlattices under study are formed by
alternating strips of gapless graphene and of its gapped
modifications. The latter can be produced using the
main property of graphene, namely, its two-dimensiona-lity.
For this, there exist two possible ways: (i)
choosing the material of the substrate on which
graphene is deposited and (ii) depositing atoms or
molecules, e.g., hydrogen atoms \citep{Elias} or CrO$_3$ molecules \citep{Zanella}
on the surface of a graphene sheet. Several
gapped modifications of graphene with the band gap
ranging from about 10~meV to 1 eV have been already
obtained.

In the superlattice under study, the charge carriers
effectively acquire a nonzero mass. Their dispersion
law becomes nonlinear. The system is similar to a relativistic
plasma in a low-dimensional space.

Plasma waves in the graphene superlattice in the
presence of a high dc electric field were recently studied
in \citep{Glazov1} in the random phase approximation. The
same authors \citep{Glazov2} studied numerically the plasmon dispersion
law in the planar graphene superlattice. In our
work, we obtain explicit analytical results for the plasmon
dispersion law in the planar graphene superlattice.

\begin{center}
2. EFFECTIVE MODEL DESCRIPTION\\
OF THE SUPERLATTICE

\textit{2.1. Fundamentals of the Model Description\\
of the Superlattice}
\end{center}

The main concepts concerning the planar superlattices
based on gapless graphene and on its gapped
modifications were reported in \citep{Ratnikov1}.

Let $x$ and $y$ axes be normal and parallel to the interfaces,
respectively (\hyperlink{fig1}{Fig. 1}). The charge carriers in a
superlattice are described by the Dirac equation
\begin{equation}\label{1}
\left(\texttt{v}_\text{F}{\boldsymbol\sigma}\widehat{\bf p}+\sigma_z\Delta+V\right)\Psi(x,\,y)=E\Psi(x,\,y),
\end{equation}
where $\texttt{v}_\text{F}\approx10^8$ cm/s is the Fermi velocity; ${\boldsymbol\sigma}=(\sigma_x,\,\sigma_y)$ and $\sigma_z$ are the Pauli matrices; and $\widehat{\bf p}=-i{\boldsymbol\nabla}$ is the~momentum operator (we use units with $\hbar=1$). The half-width $\Delta$ of the band gap and the work function $V$ are periodically modulated along the $x$ axis
\begin{equation}\label{2}
\begin{split}
\Delta&=\begin{cases} 0,& d(n-1)<x<-d_\text{II}+dn,\\
\Delta_0,& -d_\text{II}+dn<x<dn,
\end{cases}\\
V&=\begin{cases} 0,& d(n-1)<x<-d_\text{II}+dn,\\
V_0,& -d_\text{II}+dn<x<dn.
\end{cases}
\end{split}
\end{equation}
where $n$ is an integer enumerating the superlattice
supercells; $d_\text{I}$ and $d_\text{II}$ are the widths of strips of gapless
and gapped graphene, respectively; and $d=d_\text{I}+d_\text{II}$ is
the period of the superlattice (see \hyperlink{fig1}{Fig. 1}). The profile
of the potential is depicted in \hyperlink{fig2}{Fig. 2}.

\begin{figure}[t!]
\begin{center}
\hypertarget{fig1}{}
\includegraphics[width=0.5\textwidth]{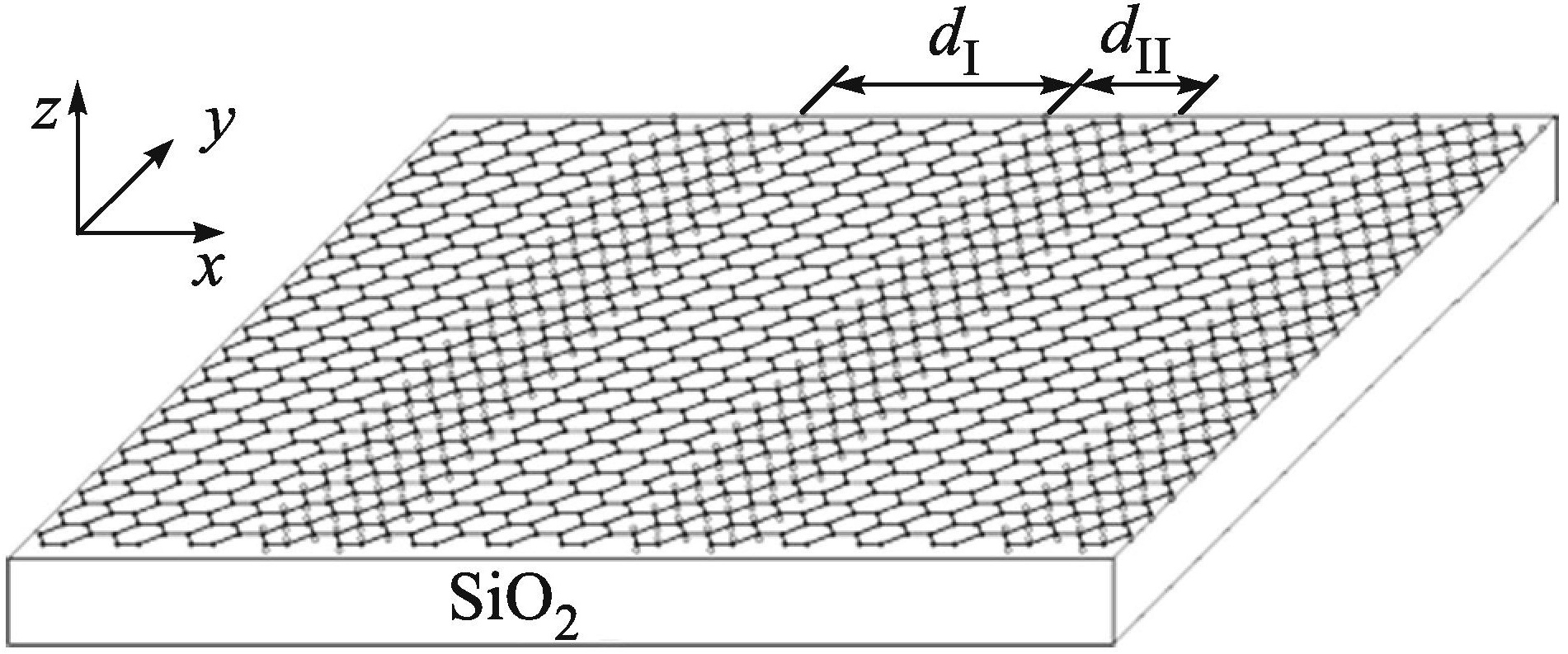}
\end{center}

\begin{list}{}
{\rightmargin=0.27cm\leftmargin=0.27cm}
\item
\footnotesize{\bf Fig. 1.} Example of an array under study: graphene sheet on
SiO$_2$ substrate with hydrogen atoms periodically deposited on graphene strips (graphene–graphane superlattice).
\end{list}
\normalsize
\end{figure}

In this work, we assume that $\texttt{v}_\text{F}$ has the same value
over the whole superlattice. In \citep{Ratnikov2}, we considered a
new type of superlattice with alternating Fermi velocity.

The motion of charge carriers along the y axis is
free and the wavefunction has the form $\Psi(x,\,y)=\psi(x)e^{ik_yy}$.

The dispersion relation for decaying solution \eqref{1}
within the potential barriers has the form \citep{Ratnikov1}
\begin{equation}\label{3}
\begin{split}
&\frac{\texttt{v}^2_\text{F}k^2_2-\texttt{v}^2_\text{F}k^2_1+V^2_0-\Delta^2_0}{2\texttt{v}^2_\text{F}k_1k_2}\sinh(k_2d_\text{II})\sin(k_1d_\text{I})\\
&+\cosh(k_2d_\text{II})\cos(k_1d_\text{I})=\cos(k_xd),
\end{split}
\end{equation}
where $k_1$ and $k_2$ are related to the energy $E$ by the formulas
\begin{equation}\label{4}
\begin{split}
E&=\pm\texttt{v}_\text{F}\sqrt{k^2_y+k^2_1},\\
E&=V_0\pm\sqrt{\Delta_0^2+\texttt{v}^2_\text{F}k^2_y-\texttt{v}^2_\text{F}k^2_2}.
\end{split}
\end{equation}

For the further analysis, it is difficult to use the
exact spectrum of charge carriers determined by finding
the numerical solution of Eq. \eqref{3}. We suggest using
the effective spectrum (the spectrum of a model narrow-gap semiconductor).

\begin{figure}[t!]
\begin{center}
\hypertarget{fig1}{}
\includegraphics[width=0.43\textwidth]{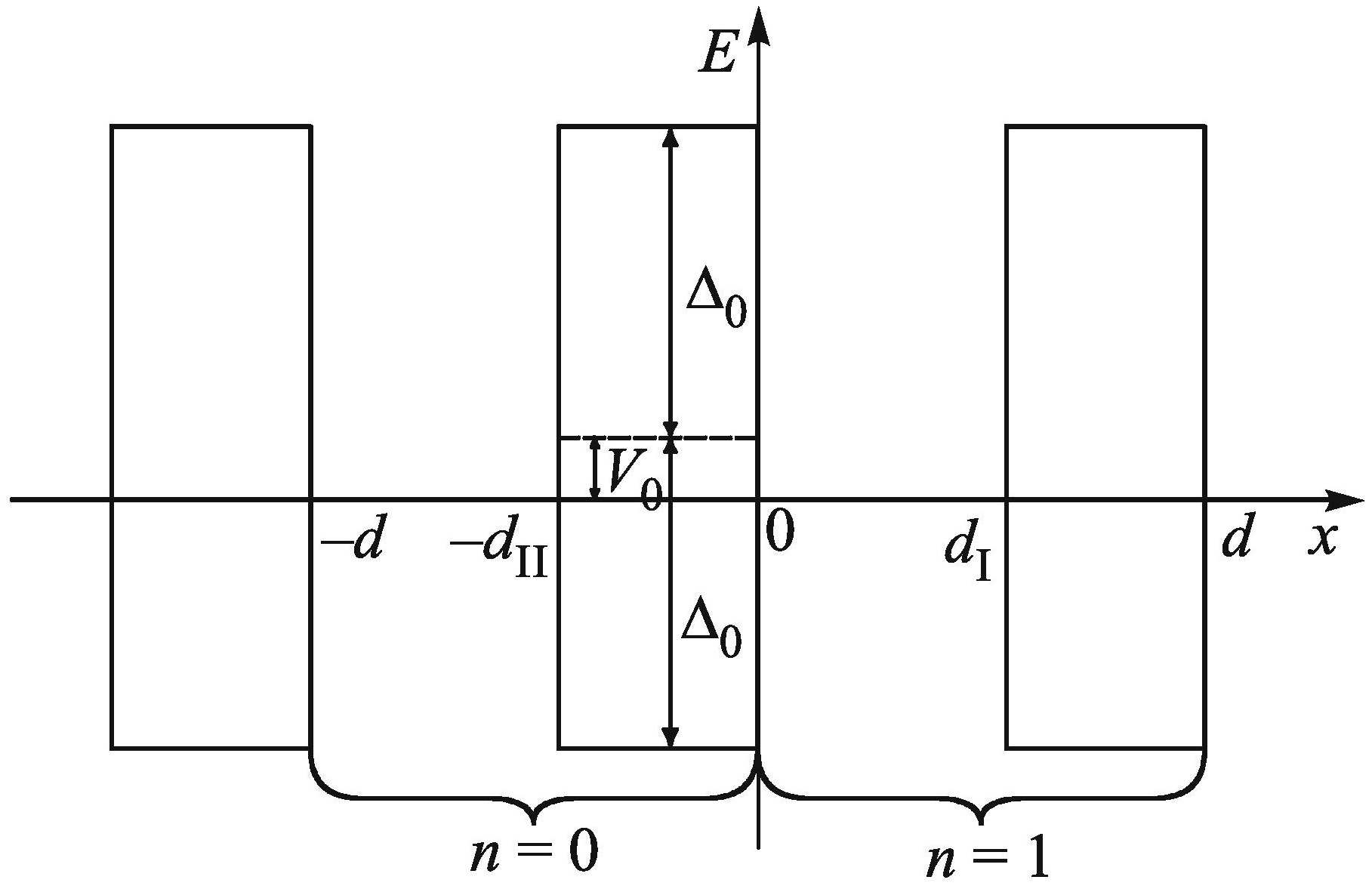}
\end{center}

\begin{list}{}
{\rightmargin=0.27cm\leftmargin=0.27cm}
\item
\footnotesize{\bf Fig. 2.} Model periodic one-dimensional Kronig–Penney
potential for the superlattice under study.
\end{list}
\normalsize
\end{figure}

\begin{center}
\textit{2.2. Effective Theory}
\end{center}

We should distinguish two cases: (i) the Fermi level
falls within one of the minigaps and (ii) the Fermi
level is located within one of the minibands.

In the former case, all minibands lying below the
Fermi level are completely occupied and the oscillations
of the electron (hole) density occur only in the
direction of the free motion of charge carriers (along
the normal to the direction of the voltage applied
across the superlattice). This is a quasi-one-dimensional
motion.

In the latter case, the miniband containing the
Fermi level is occupied only partially, whereas all
lower bands (if such bands exist) are completely occupied.
In the partially occupied miniband, the oscillations
of electron (hole) density can also occur along
the direction of the voltage applied across the superlattice.
This is a quasi-two-dimensional motion.

Then, for simplicity, we consider the situation with
the filling (complete or partial) of only one lowest
electron miniband or the highest hole miniband.

\textbf{2.2.1. Quasi-one-dimensional case (complete-ly
occupied miniband).} At sufficiently large values of $\Delta_0$
and $d_\text{II}$, the minibands are rather narrow (we shall
specify this condition below). In this case, the energy
spectrum of charge carriers is similar to that characteristic
of a quasi-one-dimensional narrow-gap semiconductor
\begin{equation}\label{5}
E\approx V_\text{eff}\pm\sqrt{\Delta^2_\text{eff}+\texttt{v}^2_\text{F}k^2_y}.
\end{equation}
The parameters $\Delta_\text{eff}$ and $V_\text{eff}$ play the role of the effective
band gap and the effective work function, respectively.
The charge carriers have the effective mass
\begin{equation}\label{6}
m^*=\frac{\Delta_\text{eff}}{\texttt{v}^2_\text{F}}.
\end{equation}

Using dispersion relation \eqref{3} and assuming that $|V_\text{eff}|<\Delta_\text{eff}\ll\Delta_0$, we can easily deduce the following estimates for $\Delta_\text{eff}$ and $V_\text{eff}$:
\begin{equation}\label{7}
\begin{split}
\Delta_\text{eff}&=\frac{\pi\texttt{v}_\text{F}}{2d_\text{I}}\left[1-\frac{\texttt{v}_\text{F}}{d_\text{I}\Delta_0}\right],\\
V_\text{eff}&=\frac{\texttt{v}_\text{F}}{d_\text{I}\Delta_0}V_0.
\end{split}
\end{equation}

In the case under study, the minibands have an
exponentially small width owing to an exponentially
small probability for charge carriers to tunnel through
the barriers. In this limit, we obtain the following estimate
for the miniband width:
\begin{equation}\label{8}
\delta E=\frac{4\texttt{v}_\text{F}}{d_\text{I}}\exp\left(-\frac{d_\text{II}}{\texttt{v}_\text{F}}\Delta_0\right).
\end{equation}
The condition defining the narrow minibands is $\delta E\ll\Delta_\text{eff}$.
Comparing the expression for $\Delta_\text{eff}$ in Eqs. \eqref{7} with
Eq. \eqref{8}, we find the condition $\Delta_0\gtrsim2\texttt{v}_\text{F}/d_\text{II}$.

Let us write the effective Hamiltonian corresponding
to the approximate dispersion law given by Eq. \eqref{5}
as the Dirac Hamiltonian in terms of 2$\times$2 matrices
\begin{equation}\label{9}
\widehat{H}^{(1D)}_\text{eff}=\texttt{v}_\text{F}\sigma_y\widehat{p}_y-\sigma_z\Delta_\text{eff}+V_\text{eff}.
\end{equation}
Here, the minus sign in front of the second term is
placed for convenience of further calculations. This~does not affect the final results since there $\Delta_\text{eff}$ is
squared.

In the zeroth order approximation, the Green’s
function describing the free propagation of charge carriers
along the gapless graphene strips has the form of
the inverse operator \citep{Kotov}
\begin{equation}\label{10}
\widehat{G}^{(1D)}_{0}(k_y,\,\omega)=\left[\omega+\mu-\widehat{H}^{(1D)}_\text{eff}\right]^{-1},
\end{equation}
where $\mu$ is the chemical potential (coincides with the
Fermi energy).

Substituting Eq. \eqref{9} into operator \eqref{10}, we can
explicitly write the Green's function taking into
account the rules of path tracing around the poles
\begin{equation}\label{11}
\begin{split}
&\widehat{G}^{(1D)}_0(k_y,\,\omega)=\frac{1}{2\varepsilon_{k_y}}\\
&\times\sum_{s=\pm1}s\frac{\omega+\widetilde{\mu}-\sigma_z\Delta_\text{eff}+\texttt{v}_\text{F}\sigma_yk_y}
{\omega+\widetilde{\mu}-s\varepsilon_{k_y}-i\delta\sgn(\widetilde{\mu}-s\varepsilon_{k_y})},
\end{split}
\end{equation}
where $\widetilde{\mu}=\mu-V_\text{eff}$ and $\varepsilon_{k_y}=\sqrt{\Delta^2_\text{eff}+\texttt{v}^2_\text{F}k^2_y}$, $\delta\rightarrow+0$.

The value of is related to the Fermi momentum $p_\text{F}$ as follows:
\begin{equation}\label{12}
|\widetilde{\mu}|=\sqrt{\Delta^2_\text{eff}+\texttt{v}^2_\text{F}p^2_\text{F}}.
\end{equation}
The one-dimensional Fermi momentum is expressed
in terms of the charge carrier density
\begin{equation}\label{13}
p_\text{F}=\frac{\pi}{g}n_{2D}d,
\end{equation}
where $g=g_sg_v$ is the degeneracy order ($g_s=2$ is the spin
degeneracy order and $g_v=2$ is the valley degeneracy
order).

\textbf{2.2.1. Quasi-two-dimensional case (partially
occupied miniband).} In the quasi-two-dimensional case, in
addition to the free motion along the gapless graphene
strips, charge carriers move across the potential barriers.
These types of motion occur at different velocities: at $\texttt{v}_\parallel$
for the free motion and at a much lower velocity $\texttt{v}_\perp\ll\texttt{v}_\parallel$
(since the probability of tunneling through the potential
barrier is small). This means the quasi-two-dimensional
anisotropic motion of charge carriers.
The corresponding values of $\texttt{v}_\perp$ and $\texttt{v}_\parallel$ are selected by
fitting the approximate dispersion law
\begin{equation}\label{14}
E\approx V_{eff}\pm\sqrt{\Delta^2_{eff}+\texttt{v}^2_\perp k^2_x+\texttt{v}^2_\parallel k^2_y}.
\end{equation}
The energy spectrum is similar to that of an anisotropic
narrow-band semiconductor with the effective
masses $m^*_\perp=\Delta_\text{eff}/\texttt{v}^2_\perp\neq m^*_\parallel=\Delta_\text{eff}/\texttt{v}^2_\parallel$. The temperature should be sufficiently low, $T\ll\delta E$.

The effective Hamiltonian with eigenvalues \eqref{14}
has the form
\begin{equation}\label{15}
\widehat{H}^{(2D)}_\text{eff}=\texttt{v}_\perp\sigma_x\widehat{p}_x+\texttt{v}_\parallel\sigma_y\widehat{p}_y-\sigma_z\Delta_\text{eff}+V_\text{eff}.
\end{equation}

The Green’s function is determined as inverse
operator \eqref{10} with the Hamiltonian
\begin{equation}\label{16}
\begin{split}
&\widehat{G}^{(2D)}_{0}({\bf k},\,\omega)=\frac{1}{2\varepsilon_{\bf k}}\\
&\times\sum_{s=\pm1}s\frac{\omega+\widetilde{\mu}-\sigma_z\Delta_\text{eff}+\texttt{v}_\perp\sigma_xk_x+\texttt{v}_\parallel\sigma_yk_y}
{\omega+\widetilde{\mu}-s\varepsilon_{\bf k}-i\delta\sgn(\widetilde{\mu}-s\varepsilon_{\bf k})},
\end{split}
\end{equation}
where $\varepsilon_{\bf k}=\sqrt{\Delta^2_\text{eff}+\texttt{v}^2_\perp k^2_x+\texttt{v}^2_\parallel k^2_y}$.

\begin{center}
3. PLASMONS

\textit{3.1. Coulomb Interaction}
\end{center}

In the quasi-one-dimensional case, the charge carriers
do not move between the gapless graphene strips.
The Coulomb interaction is similar to that for charge
carriers in a periodic planar array formed by parallel
filaments. In such array, the Coulomb interaction of
charges located at two filaments separated by the distance
$nd$ reads \citep{Andryushin}
\begin{equation}\label{17}
V(k_y,\,n)=2\widetilde{e}^2K_0\left(d|nk_y|\right),
\end{equation}
where $d$ is the distance between the gapless graphene
strips (it coincides with the period of the superlattice);
$n$ is the number of a strip (it can be considered as that
coinciding with the number of a supercell in the superlattice
shown in \hyperlink{fig2}{Fig. 2}); $\widetilde{e}^2=e^2/\varepsilon_\text{eff}$, where $\varepsilon_\text{eff}=(\varepsilon_1+\varepsilon_2)/2$ is the effective static dielectric constant determined by the static dielectric constants $\varepsilon_1$ and $\varepsilon_2$ of the
media surrounding the graphene (e.g., vacuum and
the substrate material); and $K_0(x)$ is the~modified Bessel
function of the second kind.

Now, we can make the transformation from the
discrete variable $n$ denoting the strip number to the
dimensionless transverse momentum
$\theta=k_xd$ ($-\pi\leq\theta\leq\pi$), as was done in \citep{Andryushin}
\begin{equation}\label{18}
\begin{split}
&V(k_y,\,\theta)=\sum_{n=-\infty}^\infty V(k_y,\,n)e^{in\theta}\\
\hspace{-0.12cm}=2\widetilde{e}^2&K_0\left(\frac{d_I}{2}|k_y|\right)+4\widetilde{e}^2\sum_{n=1}^\infty\cos(n\theta)K_0\left(nd|k_y|\right).
\end{split}
\end{equation}

In the case of the narrow barrier, which is of main
interest to us, expression \eqref{18} becomes simpler ($d_\text{II}\ll d_\text{I}$) \citep{Andryushin}
\begin{equation}\label{19}
\begin{split}
&V(k_y,\,\theta)=2\widetilde{e}^2\ln\frac{d}{\pi d_I}\\
&+\left[-2C-2\psi\left(\frac{\theta}{2\pi}+\frac{1}{2}\right)+\pi\tan\frac{\theta}{2}\right]\widetilde{e}^2+o(k_yd),
\end{split}
\end{equation}
where $C=0.577\ldots$ is the Euler constant and $\psi(x)$ is the
Euler $\psi$ function. At the miniband boundaries, we have
\begin{equation}\label{20}
V(k_y,\,\pm\pi)=2\widetilde{e}^2\ln\frac{d}{\pi d_\text{I}}+\frac{2\pi\widetilde{e}^2}{|k_y|d}+o(k_yd).
\end{equation}

\begin{center}
\textit{3.2. Polarization Operator}
\end{center}

In the calculations of the plasmon frequencies
using the diagram technique, we should distinguish
two specific cases: (i) the quasi-one-dimensional isotropic
case (the corresponding Green’s function is
determined in Subsection 2.2.1) and (ii) the quasi-two-dimensional
anisotropic case (the corresponding
Green's function is determined in Subsection 2.2.2).
Hence, we have two expressions for the polarization
operator needed for finding the plasmon dispersion
law.

\textbf{3.2.1. Quasi-one-dimensional polarization operator.} The polarization operator is represented by the
loop diagram (\hyperlink{fig3}{Fig. 3}) and is given by the expression
\begin{equation}\label{21}
\begin{split}
&\Pi^{(1D)}(k_y,\,\omega)=-ig\int\frac{dp_y}{2\pi}\\
\times\int\frac{d\varepsilon}{2\pi}\Tr&\left\{\widehat{G}^{(1D)}_{0}(p_y,\,\varepsilon)\widehat{G}^{(1D)}_{0}(p_y+k_y,\,\varepsilon+\omega)\right\}.
\end{split}
\end{equation}
Similar to the situation in quantum electrodynamics
(QED), expression \eqref{21} should be renormalized.
However, the many-body problem in solids has its specific
features. Although the bare electron and hole
spectra are identical to those of electrons and positrons
in QED, the set of parameters and the laws
involved in these renormalizations are different \citep{Zawadzki, Gelmont}.

The renormalization of the polarization operator is
reduced to the condition \citep{Markova}
\begin{equation}\label{22}
\Pi^{(1D)}_\text{Ren}(k_y,\,\omega)=\Pi^{(1D)}(k_y,\,\omega)-\left.
\Pi^{(1D)}(k_y,\,\omega)\right|_{n_{2D}\rightarrow0}.
\end{equation}

We are interested in plasmons (the long-wavelength
collective excitations); therefore, it is sufficient
to determine the polarization operator at low $k_y$ and $\omega$
values:
\begin{equation}\label{23}
|k_y|\ll\frac{\Delta_\text{eff}}{\texttt{v}_\text{F}},\,|\omega|\ll\Delta_\text{eff}.
\end{equation}
As we can see below, the plasmon frequencies are low
because of low $k_y$ values (the plasmon dispersion law
for low-dimensional systems).

In the quasi-one-dimensional case, the real part of
the renormalized polarization operator at low crystal
momenta (the expansion is performed up to terms of
the order of $k^2_y$) and frequencies specified by Eqs. \eqref{23}
is given by the expression
\begin{equation}\label{24}
\begin{split}
&\RE \Pi^{(1D)}_\text{Ren}(k_y,\,\omega)=\frac{g}{2\pi}\\
&\times\left\{-\Lambda_1+\frac{|k_y|}{|\widetilde{\mu}|}+\frac{2\texttt{v}^2_\text{F}p_\text{F}k^2_y}{|\widetilde{\mu}|\omega^2}+\frac{\texttt{v}^4_\text{F}p^3_\text{F}k^2_y}{3|\widetilde{\mu}|^3\Delta^2_\text{eff}}\right\},
\end{split}
\end{equation}
where
\begin{equation}\label{25}
\Lambda_1=\frac{1}{\texttt{v}_\text{F}}\ln\frac{|\widetilde{\mu}|+\texttt{v}_\text{F}p_\text{F}}{|\widetilde{\mu}|-\texttt{v}_\text{F}p_\text{F}}.
\end{equation}
Note that $\Lambda_1$ is positive, is independent of both $k_y$ and
$\omega$, and appears in Eq. \eqref{24} with the negative sign. It
easy to see that this term in Eq. \eqref{24} results in a pronounced
(background) screening. Hence, $\Lambda_1$ should
be omitted. In addition, in the limit $\Delta_\text{eff}\rightarrow0$, it leads
to a logarithmic divergence.

\begin{figure}[t!]
\begin{center}
\hypertarget{fig3}{}
\includegraphics[width=0.4\textwidth]{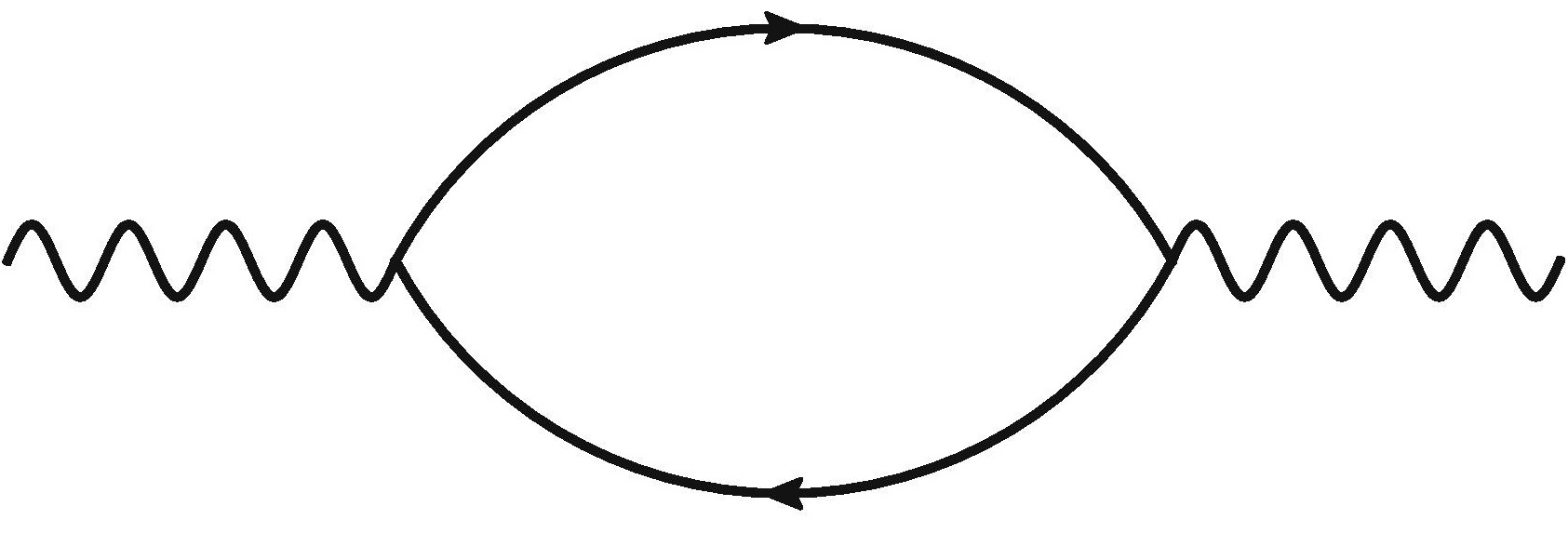}
\end{center}

\begin{list}{}
{\rightmargin=0.27cm\leftmargin=0.27cm}
\item
\begin{center}
{\footnotesize{\bf Fig. 3.} Loop diagram.}
\end{center}
\end{list}
\normalsize
\end{figure}

The term linear in $|k_y|$ actually turns out to be small
in comparison to the third term, which is formally of
the order of $k^2_y$. In the denominator of the third term,
we have $\omega^2$, and $\omega\gtrsim\texttt{v}_\text{F}|k_y|$, but it is still of the same
order as $\texttt{v}_\text{F}|k_y|$. Therefore, in spite of the formally higher order of the third term, it turns out to be larger than the second one.

Expression \eqref{24} is derived under the assumption that $\texttt{v}_\text{F}|k_y|\lesssim|\omega|\ll\Delta_\text{eff}$. At the same time, we assume that $\texttt{v}_\text{F}p_\text{F}\ll\Delta_\text{eff}$ (the case of a low charge carrier density)
or, at least, $\texttt{v}_\text{F}p_\text{F}\lesssim\Delta_\text{eff}$ (the case of a moderate
charge carrier density). Therefore, in contrast to the
third term, the last term in Eq. \eqref{24} is smaller than or
of the order of $\omega^2/\Delta^2_\text{eff}$ in the case of a moderate
charge carrier density and of a higher order in the case
of a low charge carrier density. Therefore, we can
neglect the last term in Eq. \eqref{24}.

Finally, we find
\begin{equation}\label{26}
\RE \Pi^{(1D)}_\text{Ren}(k_y,\,\omega)=\frac{g\texttt{v}^2_\text{F}p_\text{F}k^2_y}{\pi|\widetilde{\mu}|\omega^2}.
\end{equation}

The imaginary part of $\Pi^{(1D)}_\text{Ren}(k_y,\,\omega)$ vanishes
within the range
\begin{equation}\label{27}
\texttt{v}_\text{F}|k_y|<|\omega|<\sqrt{4\Delta^2_\text{eff}+\texttt{v}^2_\text{F}k^2_y},
\end{equation}
which is in agreement with the well-known result for
relativistic plasma \cite{Tsitovich}.

\textbf{3.2.2. Quasi-two-dimensional polarization operator.} In the quasi-two-dimensional anisotropic case,
the polarization operator is represented similarly to Eq. \eqref{21} as
\begin{equation}\label{28}
\begin{split}
&\Pi^{(2D)}({\bf k},\,\omega)=-igd\int\frac{d^2p}{(2\pi)^2}\\
\times\int\frac{d\varepsilon}{2\pi}\Tr&\left\{\widehat{G}^{(2D)}_{0}({\bf p},\,\varepsilon)\widehat{G}^{(2D)}_{0}({\bf p}+{\bf k},\,\varepsilon+\omega)\right\}.
\end{split}
\end{equation}

The renormalization condition in the form of
Eq. \eqref{22} should also be imposed on polarization operator
\eqref{28}. At low crystal momenta (we retain the terms
of the order of $k^2_x$ and $k^2_y$) and low frequencies specified
by Eqs. \eqref{23}, the real part of the renormalized
polarization operator has the form
\begin{equation}\label{29}
\begin{split}
&\RE\Pi^{(2D)}_\text{Ren}({\bf k},\,\omega)=\frac{gd}{2\pi}\left\{-\Lambda_2+\frac{\texttt{v}_\perp^2k^2_x+\texttt{v}_\parallel^2k^2_y}{\texttt{v}_{\perp}\texttt{v}_\parallel}\right.\\
&\left.\times\left[\frac{\widetilde{\mu}^2-\Delta^2_\text{eff}}{2|\widetilde{\mu}|\omega^2}+\frac{1}{6\Delta_\text{eff}}
\left(1-\frac{3\Delta_\text{eff}}{2|\widetilde{\mu}|}+\frac{\Delta^3_\text{eff}}{2|\widetilde{\mu}|^3}\right)\right]\right\},
\end{split}
\end{equation}
which includes the positive parameter
\begin{equation}\label{30}
\Lambda_2=\frac{|\widetilde{\mu}|-\Delta_\text{eff}}{\texttt{v}_{\perp}\texttt{v}_\parallel}.
\end{equation}
This parameter should also be omitted for the same
reasons as in the case of $\Lambda_1$ in Eq. \eqref{24}. The second
term in the square brackets is smaller than the first one
by a factor of $\omega^2/(\widetilde{\mu}^2-\Delta^2_\text{eff})$ and, therefore, can also
be omitted. Thus, we obtain
\begin{equation}\label{31}
\RE\Pi^{(2D)}_\text{Ren}({\bf k},\,\omega)=\frac{gd}{4\pi}\frac{\texttt{v}_\perp^2k^2_x+\texttt{v}_\parallel^2k^2_y}{\texttt{v}_{\perp}\texttt{v}_\parallel}\frac{\widetilde{\mu}^2-\Delta^2_\text{eff}}{|\widetilde{\mu}|\omega^2}.
\end{equation}

The imaginary part of $\Pi^{(2D)}_\text{Ren}({\bf k},\,\omega)$ vanishes within
the range
\begin{equation}\label{32}
\sqrt{\texttt{v}_\perp^2k_x^2+\texttt{v}_\parallel^2k_y^2}<|\omega|<\sqrt{4\Delta^2_\text{eff}+\texttt{v}_\perp^2k_x^2+\texttt{v}_\parallel^2k_y^2}.
\end{equation}

\begin{center}
\textit{3.3. Dispersion Law for Plasmons}
\end{center}

In the random phase approximation, the dispersion
law for plasmons is determined by the equation
\begin{equation}\label{33}
1-V({\bf k})\Pi({\bf k},\,\omega)=0.
\end{equation}
When the Fermi level falls within the minigap, Eq. \eqref{26} for the polarization operator $\Pi({\bf k},\,\omega)$ and Eq. \eqref{18} for the Coulomb interaction should be substituted
into Eq. \eqref{33}. When the Fermi level falls within the miniband, Eq. \eqref{31} for the polarization operator $\Pi({\bf k},\,\omega)$ and Eq. \eqref{18} with $\theta=k_xd$ for the Coulomb
interaction should be substituted into Eq. \eqref{33}. In the former case, we obtain
\begin{equation}\label{34}
\omega^{(1D)}_{pl}(k_y,\,\theta)=\texttt{v}_\text{F}|k_y|\sqrt{\frac{gp_\text{F}}{\pi|\widetilde{\mu}|}V(k_y,\,\theta)}.
\end{equation}
In the latter case, we have
\begin{equation}\label{35}
\omega^{(2D)}_{pl}({\bf k})=\sqrt{\texttt{v}_\perp^2k_x^2+\texttt{v}_\parallel^2k_y^2}\sqrt{\frac{gd}{4\pi}\frac{\widetilde{\mu}^2-\Delta^2_\text{eff}}{\texttt{v}_\perp \texttt{v}_\parallel|\widetilde{\mu}|}V({\bf k})}.
\end{equation}

In the case of closely spaced strips of gapless graphe-ne, expression \eqref{34} at the boundary of the plasmon band gives the square-root plasmon dispersion law characteristic of two-dimensional systems:
\begin{equation}\label{36}
\omega^{(1D)}_{pl}(k_y)=\texttt{v}_\text{F}\sqrt{\frac{2\pi n_{2D}\widetilde{e}^2}{|\widetilde{\mu}|}|k_y|}.
\end{equation}
At low $k_y$ values, we retain only the second term in
Eq. \eqref{20} for the Coulomb interaction.

However, it follows from Eq. \eqref{34} in this case that
the plasmon dispersion law remains acoustic for
nearly the whole plasmon band (almost for all $\theta$ values),
\begin{equation}\label{37}
\omega^{(1D)}_{pl}(k_y,\,\theta)=\texttt{v}_\text{F}|k_y|\sqrt{\frac{2g\widetilde{e}^2p_\text{F}}{\pi|\widetilde{\mu}|}f(\theta)},
\end{equation}
where
\begin{equation}\label{38}
f(\theta)=\ln\frac{d}{\pi d_I}-C-\psi\left(\frac{\theta}{2\pi}+\frac{1}{2}\right)+\frac{\pi}{2}\tan\frac{\theta}{2}
\end{equation}
according to Eq. \eqref{19} for the Coulomb interaction.

In the case of the linear dependence of the chemical
potential on the Fermi momentum, Eq. \eqref{36} gives
the well-known result for the plasmon dispersion law
in gapless graphene \citep{Kotov}
\begin{equation}\label{39}
\omega_{pl}(k_y)=\sqrt{\frac{g}{2}|\widetilde{\mu}|\widetilde{e}^2|k_y|}.
\end{equation}
Here, the plasmon propagates along the $y$ axis.

The dispersion law for the two-dimensional plasmon
in gapless graphene can also be obtained from
Eq. \eqref{35} in the isotropic case, where $\texttt{v}_\perp=\texttt{v}_\parallel=\texttt{v}_\text{F}$ and $\widetilde{\mu}^2-\Delta^2_{eff}=\texttt{v}^2_\text{F}p^2_\text{F}$. Here, in the quasi-two-dimensional
case, we should take into account the relation
\begin{equation}\label{40}
p^2_\text{F}=\frac{4\pi}{g}n_{2D}.
\end{equation}

Formulas \eqref{34} and \eqref{35} give the well-known
expressions for the case of nonrelativistic charge carriers
\citep{Andryushin}. For example, at large distances between the
strips of gapless graphene ($d_\text{II}\gg d_\text{I}$), the system
behaves as a set of strips. The Coulomb interaction
between the charge carriers in one of such strips is
given by the first term on the right-hand side of
Eq. \eqref{18}.

In the nonrelativistic limit, when $\texttt{v}_\text{F}p_\text{F}\ll\Delta_\text{eff}$ and
$|\widetilde{\mu}|\approx\Delta_\text{eff}$, Eq. \eqref{34} yields
\begin{equation}\label{41}
\omega^{(1D)}_{pl}(k_y)=|k_y|\sqrt{\frac{2g\widetilde{e}^2p_\text{F}}{\pi m^*}\ln\frac{4}{|k_y|d_I}}.
\end{equation}
In the nonrelativistic limit for the case of isotropy with
respect to velocities, formula \eqref{35} gives
\begin{equation}\label{42}
\omega^{(2D)}_{pl}({\bf k})=\Omega_p\sqrt{|{\bf k}|d},
\end{equation}
where
\begin{equation}\label{43}
\Omega_p=\left(\frac{2\pi\widetilde{e}^2n_{2D}}{dm^*}\right)^{1/2}.
\end{equation}

\begin{center}
\textit{3.4. Band Character of Plasmon Excitations}
\end{center}

Owing to the periodicity of the array under study,
not only the spectrum of single-particle excitations but
also the plasmon excitation spectrum is separated into
minibands. In the momentum space, the boundaries
of plasmon bands coincide with the boundaries of the
corresponding minibands for the charge carriers. This
is a consequence of the Bragg condition: $2{\bf kg}_j={\bf g}^2_j$,
$j=\pm1,\,\pm2,\,\ldots$, where ${\bf g}_j=(2\pi j/d,\,0)$ is the reciprocal
lattice vector related to the potential of the superlattice.
Thus, we find $k_{xj}=\pi j/d$.

A discontinuity appears at the boundaries of plasmon
bands. Similar to \citep{Chaplik}, we can find boundary values
for the plasmon frequencies.

The values of plasmon frequencies in the center of
the plasmon band $\omega_{pl}(0)$ in higher minibands coincide
with the minimum energy values for charge carriers in
these minibands. Let us estimate these values. Finding
an approximate solution of dispersion relation \eqref{3} with
respect to energy at the point $k_x=k_y=0$, we obtain the
following estimate for the energy of charge carriers in
the $n$th miniband ($n\,=\,0,\,1,\,2,\,\ldots$):
\begin{equation}\label{44}
E^{e,\,h}_n=\frac{\texttt{v}_\text{F}}{d_\text{I}}\left[\pm\left(\frac{\pi}{2}+\pi n\right)\left(1-\frac{\texttt{v}_\text{F}}{d_\text{I}\Delta_0}\right)+\frac{V_0}{\Delta_0}\right],
\end{equation}
where the upper and lower signs correspond to electrons
and holes, respectively. In particular, estimate \eqref{44} for $n=0$ (the lowest electron or highest hole bands) gives $E^{e,\,h}_0=\pm\Delta_\text{eff}+V_\text{eff}$, where $\Delta_\text{eff}$ and $V_\text{eff}$ are specified by Eqs. \eqref{7}. Let us take the characteristic values $d_\text{I}\simeq10$ nm, $\Delta_0\simeq1$ eV, and $V_0=0$ for simplicity of estimates. Then, we have $E^e_0\simeq80$ meV and $E^e_1\simeq240$ meV. We can see that the energy difference between the neighboring minibands far exceeds room temperature. Hence, we can neglect the thermally activated filling of higher minibands.

Let us now estimate how many additional electrons
are needed to completely fill the lowest electron miniband
and to start the filling of the next electron miniband.
The Fermi momentum $p_\text{F}$ should be such that
the chemical potential $\widetilde{\mu}$ becomes equal to $E^e_1$. Using
Eq. \eqref{12}, we obtain $\texttt{v}_\text{F}p_\text{F}=\sqrt{E^{e2}_1-E^{e2}_0}$ (owing to the
condition $V_0=0$, coincides with $\Delta_\text{eff}$). Then,
according to Eq. \eqref{13}, we can relate the found $p_\text{F}$ value
to the two-dimensional electron density, $n_{2D}=g\sqrt{E^{e2}_1-E^{e2}_0}/(\pi\texttt{v}_\text{F}d)\simeq5\times10^{12}$ cm$^{-2}$. This value is
fairly large as compared to the experimental data for
gapless graphene \citep{Kotov}.

\begin{center}
\textit{3.5. Absorption Intensity for the Modulated\\
Electromagnetic Radiation}
\end{center}

The intensity of absorption for electromagnetic
wa-ves modulated with the period equal to the plasmon
wavelength is given by the well-known formula
\begin{equation}\label{45}
Q=\frac{1}{2}\RE\left(\sigma\widetilde{\boldsymbol{\mathcal{E}}}\boldsymbol{\mathcal{E}}^*\right),
\end{equation}
where $\sigma$ is the conductivity of the system, $\widetilde{\boldsymbol{\mathcal{E}}}$ is the electric
field of the plasmon wave, and $\boldsymbol{\mathcal{E}}$ is the electric
field of the electromagnetic wave varying with the~frequency~$\omega$.

The conductivity of the system is easily found from
the kinetic equation in the $\tau$ approximation. In the
quasi-one-dimensional case, it reads
\begin{equation}\label{46}
\sigma^{(1D)}=\frac{ige^2\texttt{v}_\text{F}p_\text{F}}{\pi|\widetilde{\mu}|(\omega+i\nu)}.
\end{equation}
Here and further on, $\nu=1/\tau$.

In the quasi-two-dimensional
case, the conductivity is a tensor with diagonal elements
\begin{equation}\label{47}
\begin{split}
\sigma^{(2D)}_{xx}=\frac{ige^2}{4\pi(\omega+i\nu)}\frac{\widetilde{\mu}^2-\Delta^2_\text{eff}}{|\widetilde{\mu}|}\frac{\texttt{v}_\perp}{\texttt{v}_\parallel},\\
\sigma^{(2D)}_{yy}=\frac{ige^2}{4\pi(\omega+i\nu)}\frac{\widetilde{\mu}^2-\Delta^2_\text{eff}}{|\widetilde{\mu}|}\frac{\texttt{v}_\parallel}{\texttt{v}_\perp}.
\end{split}
\end{equation}

\newpage
In the quasi-one-dimensional case, the absorption intensity is
\begin{equation}\label{48}
Q^{(1D)}\simeq\sigma^{(1D)}_0\mathcal{E}^2_0\frac{\omega^2\nu^2}{(\omega^2-\omega_0^2)^2+\omega^2\nu^2},
\end{equation}
where $\sigma^{(1D)}_0$ is the value of conductivity \eqref{46} in the
zero-frequency limit, $\omega_0=\omega^{(1D)}_{pl}(k_0)$ is the plasmon
frequency corresponding to the wave vector ${\bf k}_0 = (0,\,
k_0)$, and $\mathcal{E}_0$ is the electric field amplitude.

In the quasi-two-dimensional case, the anisotropy
of the conductivity leads to the dependence of absorption
on the orientation of the polarization plane in the
incident electromagnetic wave. Let its polarization
plane be rotated by the angle $\varphi$ with respect to the $x$
axis. Then, the absorption intensity is equal to
\begin{equation}\label{49}
Q^{(2D)}=Q^{(2D)}_\perp\cos^2\varphi+Q^{(2D)}_\parallel\sin^2\varphi.
\end{equation}
Here,
\begin{equation}\label{50}
\begin{split}
Q^{(2D)}_\perp&\simeq\frac{1}{2}\sigma^{(2D)}_{xx0}\mathcal{E}^2_0\frac{\omega^2\nu^2}{(\omega^2-\omega_0^2)^2+\omega^2\nu^2},\\
Q^{(2D)}_\parallel&\simeq\frac{1}{2}\sigma^{(2D)}_{yy0}\mathcal{E}^2_0\frac{\omega^2\nu^2}{(\omega^2-\omega_0^2)^2+\omega^2\nu^2},
\end{split}
\end{equation}
where $\sigma^{(2D)}_{xx0}$ and $\sigma^{(2D)}_{yy0}$ are the expressions for the diagonal
elements of conductivity tensor \eqref{47} in the zero-frequency
limit and $\omega_0=\omega^{(2D)}_{pl}({\bf k}_0)$ is the plasmon frequency
corresponding to wave vector ${\bf k}_0=(k_0\cos\varphi,$ $k_0\sin\varphi)$.

It follows from Eqs. \eqref{50} that the spectrum of
absorption of electromagnetic waves by plasmons
should be strongly anisotropic: the corresponding
ratio of intensities is $Q^{(2D)}_\perp/Q^{(2D)}_\parallel=(\texttt{v}_\perp/\texttt{v}_\parallel)^2$. Such
strong aniso-tropy could be easily revealed in an experiment.

\begin{center}
4. CONCLUSIONS
\end{center}

In this work, we have explicitly derived the plasmon
dispersion law for the planar graphene superlattice.
It~has~been shown that the absorption spectrum
for linearly polarized electromagnetic waves modulated
with the period equal to the plasmon wavelength
should exhibit a pronounced anisotropy. When the
Fermi level falls within the minigap, this anisotropy is
due to the exclusion of oscillations of the charge carrier
density across the superlattice potential. When the
Fermi level falls into the miniband, this anisotropy is
attributed to an appreciable suppression ($\sim(\texttt{v}_\perp/\texttt{v}_\parallel)^2$)
of such oscillations because of a significant difference
between the transverse and longitudinal velocity components.

We are grateful to D.\,N. Sob'yanin for helpful discussions
and valuable comments.

\begin{flushright}
\emph{Translated by K. Kugel}
\end{flushright}
\end{document}